\documentclass[final,5p,times,twocolumn,numbers,sort&compress]{elsarticle}
\usepackage[T1]{fontenc}
\usepackage{amsmath,amssymb,graphicx,booktabs,microtype,xcolor,array}
\usepackage[colorlinks=true,linkcolor=blue,citecolor=blue,urlcolor=blue]{hyperref}

\journal{Physics Letters B}

\newcommand{\GeV}{\mathrm{GeV}}

\newcommand{\keV}{\mathrm{keV}}
\newcommand{\dd}{\mathrm{d}}
\newcommand{\sig}{\sigma_{\rm ref}}
\newcommand{\ER}{E_R}
\newcommand{\Eth}{E_{1,\rm th}}
\newcommand{\Dbar}{\overline D}
\newcommand{\Jbar}{\overline J}
\begin{document}
\begin{frontmatter}
\title{Endothermic Dark Matter at LZ from a Decaying Parent}
\author[1]{Yongsoo Jho}
\ead{1jys34@gmail.com}
\author[1]{Sanghwan Kim}
\ead{sanghwankim97@yonsei.ac.kr}
\author[1,2]{Seong Chan Park}
\ead{sc.park@yonsei.ac.kr}
\affiliation[1]{organization={Department of Physics and IPAP, Yonsei University},
 city={Seoul},postcode={03722},country={Republic of Korea}}
\affiliation[2]{organization={School of Physics, Korea Institute for Advanced Study},
 city={Seoul},postcode={02455},country={Republic of Korea}}
\begin{abstract}
The LUX-ZEPLIN experiment has reported a nuclear-recoil candidate near
248 keV. We investigate a decay source for an endothermic interpretation:
a long-lived dark parent produces energetic ground-state particles that
upscatter on xenon. The parent mass fixes the injection energy, while its
lifetime determines the absolute flux through the Galactic dark-matter
column density. We include the Doppler broadening from parent and observer
motion and use exact scattering kinematics with natural xenon isotopes.
For a scalar contact interaction, a benchmark with a 10 MeV incident state
and a 506 MeV parent gives an efficiency-weighted recoil spectrum peaked
at 248.5 keV. One event in a specified high-energy window requires
$\tau_X\simeq4.4\times10^{20}\,\mathrm{s}\,
f_X[\sig/(10^{-36}\,\mathrm{cm^2})]$, with a normalization sensitive to
the nuclear response. We quantify the accompanying elastic-scattering
requirement and the target-dependent thresholds. The calculation identifies
a decay realization of the boosted flux and the portal conditions needed
to preserve the recoil signal; a detector-level fit and ultraviolet-specific
constraints remain necessary for a complete experimental assessment.
\end{abstract}
\begin{keyword}
Dark matter \sep Direct detection \sep Endothermic scattering \sep Decaying dark matter
\end{keyword}
\end{frontmatter}

\section{Introduction}
\label{sec:intro}
LUX-ZEPLIN (LZ) has reported a nuclear-recoil candidate at
$\ER=248\pm23\,(\mathrm{stat})\pm23\,(\mathrm{sys})\,\keV$ in an exposure
of 2.84 tonne-years. Across the signal models tested by the collaboration,
the background-only hypothesis has a global tension of
$2.6\sigma$~\cite{LZ2026}. The high recoil energy and the absence of a
corresponding low-energy excess motivate mechanisms that produce a hard
or localized recoil spectrum. A single event does not establish a
dark-matter (DM) signal, but it provides a useful setting in which to test
the connection between a dark-sector source and nuclear-recoil kinematics.

The candidate has prompted many interpretations. Inelastic
DM~\cite{SmithWeiner,Graham,Barello} has been studied in electroweak,
dark-photon, scalar and other portal
models~\cite{DiMauro2026,deLima2026,BaerBarger2026,LZth260926570,LZth260923691,LZth260922739,LZth260922063,LZth260921011,LZth260918564,LZth260917196,LZth260917412,LZth260915027,LZth260915413,LZth260915600,LZth260915742,LZth260915714,LZth260913038,LZth260910491,LZth260910453,LZth260910827,LZth260908893,LZth260908993,LZth260909015,LZth260909138,LZth260909385,LZth260906909,LZth260907138,LZth260907225,LZth260907451,LZth260907800,LZth260907811,LZth260906571,LZth260906825,LZth260906171,LZth260904144,LZth260904186,LZth260904163,LZth260902807,LZth260902868,LZth260901475,LZth260901583,LZth260901590,LZth260901504,LZth260901892}.
Other proposals invoke boosted particles or neutrino-induced
recoils~\cite{Kannike2026,Liang2026,LZth260924982,LZth260911600,LZth260910504,LZth260904185},
composite or extra-dimensional
DM~\cite{Jung2026,LZth260924988,LZth260923477,LZth260925114,LZth260923096,LZth260915118,LZth260915933,LZth260912045,LZth260909037,LZth260909107,LZth260909136},
and nuclear-response effects~\cite{Khan2026}. Related work discusses
electroweak scalar structure~\cite{Nomura2026} and fermionic
absorption~\cite{LouLu2026}. The implications for sidebands,
solar capture, halo modelling and complementary searches have also been
examined~\cite{Mahapatra2026,LZth260926698,LZth260921823,LZth260916529,LZth260915321,LZth260915634,LZth260915985,LZth260919174,LZth260911833,LZth260909830,LZth260910636,LZth260908712,LZth260907807,LZth260906640,LZth260906760,LZth260906750,LZth260904673,LZth260905291,LZth260904181,LZth260904175,LZth260902775,LZth260921444}.

Boosted DM can be generated by annihilation, decay or collisions with
energetic Standard-Model particles~\cite{Agashe,Bhattacharya2015,BringmannPospelov,Jho2020,Jho2021}.
Combining such production with an endothermic transition gives inelastic
boosted DM~\cite{Giudice2018,Heurtier2019}. For LZ, Alhazmi
et al.~\cite{LZth260906890} showed that a nearly monochromatic incident
flux can produce localized recoils close to an endothermic threshold.
Their analysis identifies the required flux--cross-section product and
notes that canonical Galactic annihilation falls short for their
benchmarks. Long-lived parent decays are mentioned as a possible source,
while a quantitative realization is left open.

We study that source explicitly:
\begin{equation}
 X\to\chi_1\chi_1,\qquad \chi_1 A\to\chi_2 A,
 \qquad m_2>m_1.
 \label{eq:chain}
\end{equation}
The incident energy follows from $m_X$, and the flux follows from
$f_X/\tau_X$ and the Galactic column density. Unlike an accumulated cold
excited population~\cite{Graham,BatellPospelovRitz,LZth260917935}, the
fast daughters considered here leave the Galaxy and must be
treated as a flux. We calculate its energy distribution, normalize the
recoil rate to one event, and quantify the conditions on companion elastic
scattering and excited-state decays. The result is a source calculation
within a nucleon-level effective theory; it does not assume that an
arbitrary mediator realization satisfies accelerator or cosmological
constraints. We use a real-scalar realization for the numerical benchmark.
The recoil kinematics is independent of the particle spin; an alternative
Dirac realization and its interaction-dependent normalization are collected
in \ref{app:dirac}.

\section{The decay source}
\label{sec:source}
Take $X$ to be a cold, long-lived real scalar whose present density is
$\rho_X=f_X\rho_{\rm DM}$. With
\begin{equation}
 \mathcal L\supset-\frac{g_X}{2}X\chi_1^2,
 \qquad
 \Gamma_X=\frac{g_X^2}{32\pi m_X}
 \sqrt{1-\frac{4m_1^2}{m_X^2}},
 \label{eq:parent}
\end{equation}
the parent-rest-frame energy and momentum are
$E_0=m_X/2$ and $p_0=\sqrt{E_0^2-m_1^2}$. We assume a branching fraction
of unity into this channel. A smaller branching fraction multiplies
$f_X$ in the following source expressions. The small $g_X$ can be
associated with an approximate $X\to-X$ symmetry. This explains a long
lifetime but does not fix the mass relation required for scattering near
threshold.

For ballistic daughters and negligible attenuation, the Galactic flux
integrated over directions and energies is
\begin{align}
 \Phi_G&=\frac{2f_X\Dbar}{m_X\tau_X},
 &\Dbar&=\frac{1}{4\pi}\int\dd\Omega\,D(\hat n),\nonumber\\
 D(\hat n)&=\int_0^{s_{\max}}\dd s\,
 \rho_{\rm DM}[r(s,\hat n)].
 \label{eq:flux}
\end{align}
There is no extra factor of the daughter speed in this expression: the
steady number density is inversely proportional to the transit speed.
In the zero-velocity limit,
$\dd\Phi_G/\dd E_1=\Phi_G\delta(E_1-E_0)$; the delta function supplies
the inverse-energy unit. The density in $D$ is the total DM density,
so $f_X$ appears only once.

We use an NFW profile~\cite{NFW}, with $r_s=20$ kpc,
$R_\odot=8.2$ kpc, $R_{\rm vir}=200$ kpc and
$\rho_{\rm DM}(R_\odot)=0.3\,\GeV\,\mathrm{cm}^{-3}$. It gives
\begin{equation}
 \Dbar=1.61\times10^{22}\,\GeV\,\mathrm{cm}^{-2}.
 \label{eq:Dvalue}
\end{equation}
For comparison, self-conjugate particles of mass $M$ annihilating into two
daughters give
\begin{equation}
 \Phi_{\rm ann}=\frac{f_X^2\langle\sigma v\rangle}{M^2}\Jbar,
 \qquad
 \Jbar=\frac{1}{4\pi}\int\dd\Omega\int\dd s\,\rho_{\rm DM}^2,
 \label{eq:ann}
\end{equation}
with $\Jbar=6.72\times10^{21}\,\GeV^2\,\mathrm{cm}^{-5}$ for this smooth
halo. Equal injection energies correspond to $M=E_0=m_X/2$.
Decay and annihilation can therefore give the same leading two-body
energy while having different density dependence and normalization.
Table~\ref{tab:sources} compares three production mechanisms.

\begin{table*}[t]
\centering\small\setlength{\tabcolsep}{4pt}
\begin{tabular}{p{0.15\textwidth}p{0.26\textwidth}p{0.25\textwidth}p{0.26\textwidth}}
\toprule
 & Dark-sector annihilation & Cosmic-ray scattering & Parent decay \\
\midrule
Production & $XX\to\chi_1\chi_1$ & ${\rm CR}+\chi\to{\rm CR}+\chi$ & $X\to\chi_1\chi_1$ \\
Injection energy & $E_1\simeq M$ for cold parents & Continuum set by CR spectrum and scattering & $E_0=m_X/2$ in the parent rest frame \\
Source density & $\rho_X^2\langle\sigma v\rangle/M^2$ & $\rho_\chi$ times CR flux and production cross section & $\rho_X/(m_X\tau_X)$ \\
Fraction dependence & $f_X^2$ & $f_\chi$ at fixed CR population & $f_X$ \\
For a narrow recoil band & Injection near an endothermic threshold & Requires suitable spectral or interaction structure & Injection near threshold, with parent Doppler broadening \\
Principal normalization & Annihilation rate and $J$ factor & CR distribution and production/propagation & Parent lifetime and $D$ factor \\
\bottomrule
\end{tabular}
\caption{Representative boosted-DM sources~\cite{Agashe,Bhattacharya2015,BringmannPospelov,Jho2020,Jho2021}.
The source scalings are schematic; Eq.~\eqref{eq:ann} specifies the
annihilation convention used here. A broad incident spectrum does not
exclude a CR interpretation with a suitable interaction, as illustrated
in Refs.~\cite{LZth260924982,LZth260911600}.}
\label{tab:sources}
\end{table*}

\section{Endothermic kinematics and line broadening}
\label{sec:kin}
For a nucleus of mass $M_A$ at rest, define
$s=m_1^2+M_A^2+2M_AE_1$ and
$p_f=\lambda^{1/2}(s,m_2^2,M_A^2)/(2\sqrt{s})$, where
$\lambda(a,b,c)=a^2+b^2+c^2-2ab-2ac-2bc$. The exact recoil endpoints are
\begin{align}
 E_R^\pm&=C\pm H,
 &H&=\frac{p_1p_f}{\sqrt{s}},\nonumber\\
 C&=\gamma_{\rm cm}\sqrt{M_A^2+p_f^2}-M_A,
 &\gamma_{\rm cm}&=\frac{E_1+M_A}{\sqrt{s}}.
 \label{eq:endpoints}
\end{align}
The channel opens at
\begin{equation}
 \Eth=m_2+\frac{m_2^2-m_1^2}{2M_A},
 \qquad
 E_{R,\rm th}=\frac{m_2^2-m_1^2}{2(M_A+m_2)}.
 \label{eq:threshold}
\end{equation}
These endpoints and thresholds are fixed by energy--momentum
conservation and hold for either scalar or fermionic dark states.
At the exact threshold $p_f=0$ and the scattering phase space vanishes.
For a small positive $\delta E=E_1-\Eth$,
\begin{equation}
 H\simeq\frac{M_Ap_{\rm th}}{(M_A+m_2)^2}
 \sqrt{2m_2\delta E}.
 \label{eq:width}
\end{equation}
Consequently both the injection energy and its physical spread matter.
A parent at rest fixes $E_0$, but a halo of parents does not supply a
delta-function line in the detector frame.

It is useful to distinguish the energy released in the parent decay,
$\epsilon_{\rm dec}=m_X-2m_1$, from the excess incident energy above the
scattering threshold, $\Delta E_0=E_0-\Eth$. With
$\delta=m_2-m_1>0$ they obey
\begin{equation}
 \Delta E_0=\frac{\epsilon_{\rm dec}}{2}-\delta
 -\frac{2m_1\delta+\delta^2}{2M_A}.
 \label{eq:twosplittings}
\end{equation}
Thus $\Delta E_0=0$ can still yield recoils after physical broadening,
whereas $\epsilon_{\rm dec}=0$ gives neither a decay kick nor a nonzero
two-body decay width. In the nonrelativistic limit,
$\epsilon_{\rm dec,th}\simeq2\delta(1+m_1/M_A)$ and
$v_{\rm th}^2\simeq2\delta/\mu_A$, with
$\mu_A=m_1M_A/(m_1+M_A)$. These relations connect relativistic daughters,
nonrelativistic but unbound daughters, and halo-speed daughters.
The last regime requires Galactic orbital transport when daughters can
remain bound; Eq.~\eqref{eq:flux} is used here for fast, unbound particles.
The $1/v^2$ factor in a differential cross section does not make the exact
threshold a rate maximum: the shrinking recoil interval makes the total
cross section vanish there for a regular contact amplitude.

Let $f_h(\mathbf u)$ be the normalized parent velocity distribution and
$\hat n$ point from the observer to the source. To first order in the
nonrelativistic parent and observer velocities,
\begin{equation}
 E_1=E_0+p_0\hat n\cdot(\mathbf v_E-\mathbf u),
 \label{eq:doppler}
\end{equation}
in units with $c=1$. The normalized Galactic energy distribution is
\begin{align}
 \mathcal F_G(E_1)=&\frac{1}{4\pi\Dbar}\int\dd\Omega\,D(\hat n)
 \int\dd^3u\,f_h(\mathbf u)\nonumber\\
 &\times\delta\!\left[E_1-E_0-p_0\hat n\cdot(\mathbf v_E-\mathbf u)\right].
 \label{eq:kernel}
\end{align}
We adopt $f_h\propto\exp(-u^2/v_0^2)\Theta(v_{\rm esc}-u)$ with
$v_0=220$ and $v_{\rm esc}=544$ km/s, independent of radius, and
$v_E=232$ km/s tangential to the Galactic-centre direction; the value
$v_0=238$ km/s recommended in Ref.~\cite{Baxter2021} changes the line
width below at the few-per-cent level.
Equation~\eqref{eq:kernel} includes the correlation between source
direction and observer Doppler shift. Small boost corrections to the
flux weight are neglected. The scattering kinematics itself remains exact.

\begin{figure*}[t]
\centering\includegraphics[width=\textwidth]{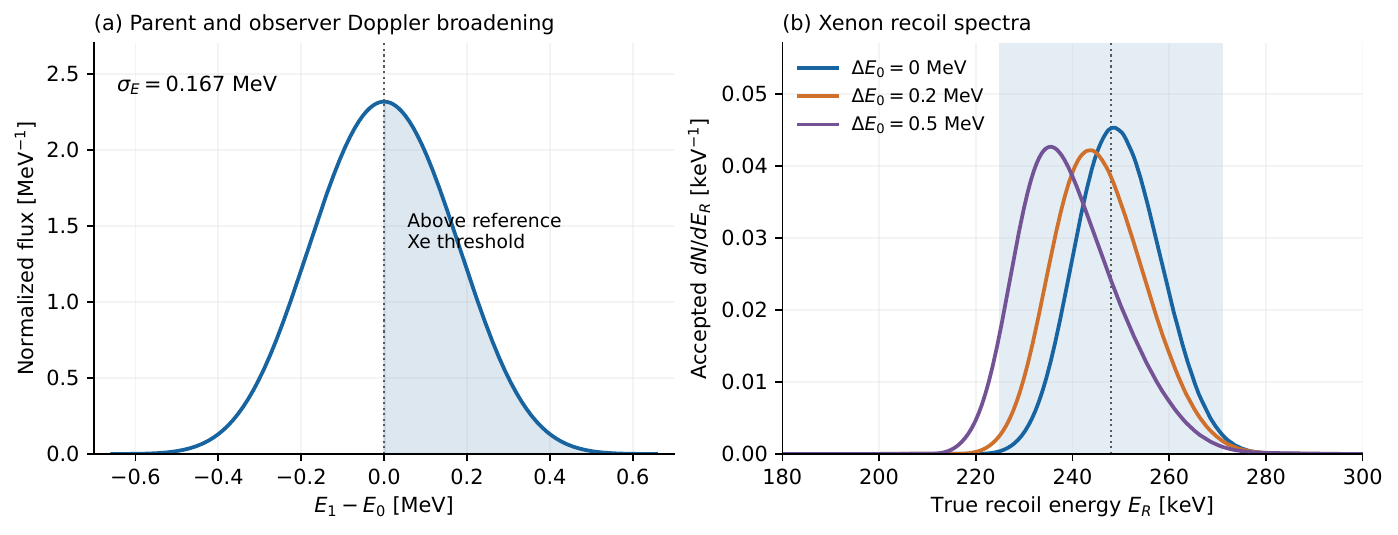}
\caption{Left: the Galactic incident-energy distribution of the benchmark
in Table~\ref{tab:bench}, including parent and observer motion. Shading
marks energies above the threshold of the reference nucleus
$M_A=122.3$ GeV; each isotope has its own threshold. Right: natural-Xe
spectra weighted by the LZ efficiency, for three shifts
$\Delta E_0=E_0-\Eth$ at fixed $(m_1,m_2)$. Each curve is normalized to
one event in the shaded true-recoil interval $225$--$271$ keV. The dotted
line marks 248 keV. Detector energy migration is not included.}
\label{fig:spectra}
\end{figure*}

For the scalar benchmark below, the standard deviation of $\mathcal F_G$
is $\sigma_E=0.167$ MeV. Its characteristic second-order Doppler
correction is below 1 keV. The finite line width allows the upper part
of the distribution to scatter even when $E_0$ equals the threshold of a
reference Xe nucleus. Figure~\ref{fig:spectra} shows the resulting flux
and recoil spectra.

The source calculation also clarifies the distinction between three
high-recoil mechanisms. Ordinary halo endothermic scattering is supplied
by the fast tail; decay-produced endothermic scattering is supplied by an
energetic ground-state flux; exothermic scattering releases the internal
energy of an incident excited state. For example, a 1 TeV halo particle
with $m_2-m_1=300$ keV requires $v_{\min}(248\,\keV)\simeq704$ km/s.
Adding a decay kick of this order does not guarantee a narrow spectrum:
the parent velocity distribution must still be folded in. At the opposite
extreme, a cold exothermic line has a different target relation,
$E_R=(m_1^2-m_2^2)/[2(M_A+m_1)]$ for $m_1>m_2$
~\cite{Graham,deLima2026,BaerBarger2026,LZth260915782}.

\section{Recoil rate and source normalization}
\label{sec:rate}
For a definite nuclear response, we use real scalar states with the
off-diagonal nucleon interaction
\begin{equation}
 \mathcal L_{\rm eff}\supset
 -C_N\chi_1\chi_2(\bar pp+\bar nn),\qquad
 \sig=\frac{C_N^2\mu_n^2}{4\pi m_1^2}.
 \label{eq:effective}
\end{equation}
Here $C_N$ has dimension $-1$, and $\mu_n$ is the $\chi_1$--nucleon
reduced mass. A scalar mediator gives $C_N=g_{12}g_N/m_\phi^2$ in the
contact limit. The reference cross section $\sig$ defines a coupling
normalization; a zero-speed endothermic process is kinematically closed.
For coherent scattering,
\begin{equation}
 \frac{\dd\sigma_A}{\dd\ER}
 =\sig\frac{A^2M_A m_1^2}{2\mu_n^2p_1^2}
 \left(1+\frac{\ER}{2M_A}\right)F_A^2(\ER)
 \label{eq:dsigma}
\end{equation}
inside the endpoints in Eq.~\eqref{eq:endpoints}, and zero outside.
We use natural isotope abundances~\cite{NIST} and the Helm
form factor~\cite{LewinSmith}. Explicitly,
$F_A=3j_1(qR_1)/(qR_1)\exp[-q^2s_H^2/2]$, with
$R_1^2=c_A^2+7\pi^2a^2/3-5s_H^2$, $c_A=1.23A^{1/3}-0.60$ fm,
$a=0.52$ fm and $s_H=0.9$ fm.

For exposure $\mathcal E$ in kg\,s, $N_T$ nuclei per kg and isotope
number fractions $\eta_A$, the accepted differential count is
\begin{equation}
 \frac{\dd N}{\dd\ER}=\mathcal E N_T\,
 \varepsilon_{\rm LZ}(\ER)\,\Phi_G
 \sum_A\eta_A\int\dd E_1\,\mathcal F_G(E_1)
 \frac{\dd\sigma_A}{\dd\ER}.
 \label{eq:rate}
\end{equation}
We use the total efficiency curve extracted from Fig.~S2 of
Ref.~\cite{LZ2026} and $\mathcal E=2.84$ tonne-years. Define
\begin{equation}
 N_W=\int_{225\,\keV}^{271\,\keV}\dd\ER\,\frac{\dd N}{\dd\ER},
 \qquad R_{\rm low}=\frac{N(14\text{--}225\,\keV)}{N_W}.
 \label{eq:windows}
\end{equation}
These are true-recoil screening quantities with efficiency weighting,
not an official LZ likelihood or reconstructed-energy selection. The
window edges coincide with the $\pm1\sigma$ statistical interval of the
candidate, a screening choice, and the lower edge of $R_{\rm low}$ (14 keV) corresponds to the onset of the high-efficiency plateau of the LZ nuclear-recoil acceptance. We do not use the
candidate's quoted errors as a detector resolution, and do not interpret
the region above the published search as a zero-event sideband.

\begin{table}[t]
\centering\small\setlength{\tabcolsep}{4pt}
\begin{tabular}{lr}
\toprule
Quantity & Benchmark \\
\midrule
$m_1$ [MeV] & 10 \\
$m_2$ [MeV] & 252.6408 \\
$m_X$ [MeV] & 505.8027 \\
$E_0$ [MeV] & 252.9013 \\
$E_{1,\text{th}}$ [MeV] & 252.9013 \\
$\sigma_E$ [MeV] & 0.167 \\
Accepted true-recoil mode [keV] & 248.5 \\
Central 90\% true-recoil interval [keV] & 235.8--264.4 \\
$R_{\rm low}$ & $6.1\times10^{-4}$ \\
$\Phi_G\sig$ for $N_W=1$ [s$^{-1}$] & $1.46\times10^{-34}$ \\
\midrule
$\sig$ [cm$^2$]; $f_X$ & $10^{-36}$; 1 \\
$C_N$ [GeV$^{-1}$] & $1.82\times10^{-4}$ \\
$\Phi_G$ [cm$^{-2}$s$^{-1}$] & 146 \\
$\tau_X$ [s] & $4.37\times10^{20}$ \\
$g_X$ [GeV] & $2.77\times10^{-22}$ \\
\bottomrule
\end{tabular}
\caption{A spectral and source-normalization benchmark for the effective
interaction in Eq.~\eqref{eq:effective}. The upper block determines the
spectrum and the required flux--coupling product. The lower block chooses
one point on the normalization relation. It is not a UV-portal exclusion
recast. The reference Xe threshold recoil is 260 keV; isotope and nuclear
weighting shift the accepted maximum to 248.5 keV.}
\label{tab:bench}
\end{table}

Table~\ref{tab:bench} demonstrates that a physical decay line can remain
localized after halo broadening. The required event count gives
\begin{align}
 \tau_X\simeq&\ 4.37\times10^{20}\,\mathrm{s}\, f_X
 \left(\frac{\sig}{10^{-36}\,\mathrm{cm}^2}\right)
 \left(\frac{\Dbar}{1.61\times10^{22}\,\GeV\,\mathrm{cm}^{-2}}\right),
 \label{eq:normalization}
\end{align}
at the benchmark masses and $N_W=1$. This relation is the main source
result. It specifies how a very small parent decay rate can supply the
required energetic population. Defining the efficiency-weighted cross
section into the window by $N_W=\mathcal E N_T\Phi_G\sigma_W$, the
benchmark gives $\sigma_W\simeq1.7\times10^{-2}\sig$ and
$\Phi_G\sigma_W\simeq2.4\times10^{-36}\,\mathrm{s^{-1}}$, which coincides
with the source-independent target of Ref.~\cite{LZth260906890}.
Table~\ref{tab:bench} is also close to the light-daughter mass point of
that work, $(m_1,m_2)\simeq(10,247)$ MeV, which places the threshold
recoil itself at 248 keV; what is added here is the source, its physical
line width and the lifetime normalization that follows from it.

At the same mean injection energy, Eq.~\eqref{eq:ann} with
$\langle\sigma v\rangle=3\times10^{-26}\,\mathrm{cm^3s^{-1}}$ and
$f_X=1$ gives $\Phi_{\rm ann}=3.15\times10^{-3}\,
\mathrm{cm^{-2}s^{-1}}$. The benchmark decay flux is larger by
$4.6\times10^4$. This is a smooth-halo source comparison at a specified
annihilation rate, not a bound on enhanced-annihilation models. A full
annihilation spectrum would also use the pair centre-of-mass velocity
distribution and $J$ weighting. Decay supplies the required normalization
through its lifetime rather than through an enhancement of that reference
annihilation rate.

\begin{figure*}[t]
\centering\includegraphics[width=\textwidth]{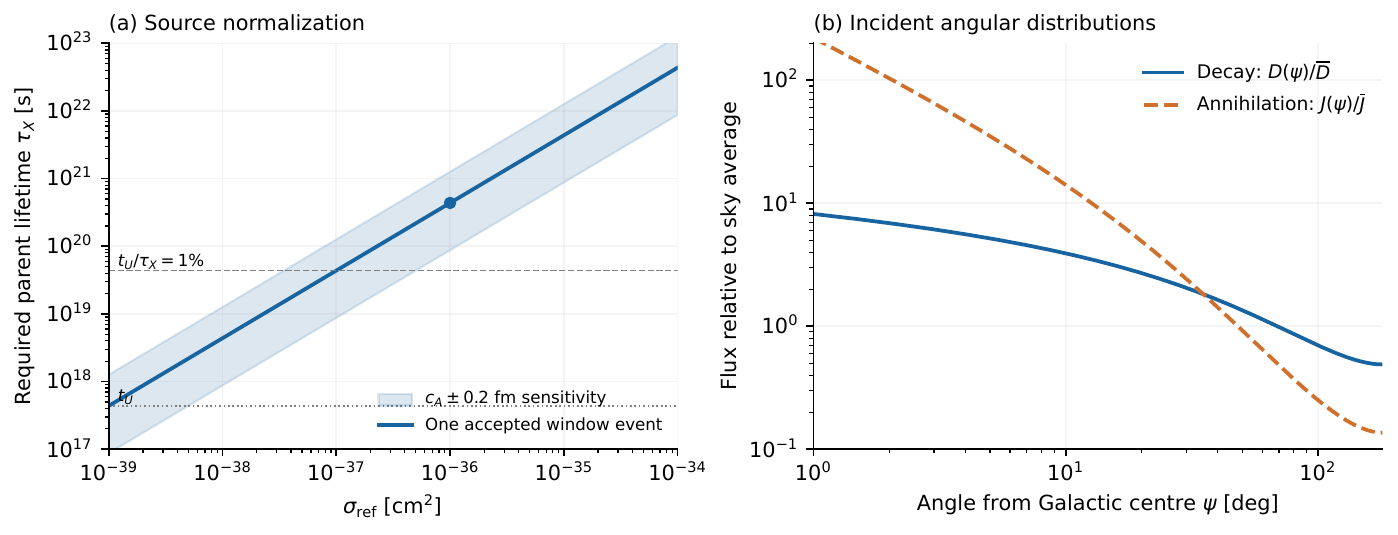}
\caption{Left: parent lifetime required for $N_W=1$ at the benchmark
masses and $f_X=1$. The band varies the Helm parameter $c_A$ by
$\pm0.2$ fm; it is not an experimental confidence region. Horizontal
lines indicate the age of the Universe and an illustrative 1\% decay
fraction; the CMB limit on decays into dark radiation,
$\tau_X\gtrsim5\times10^{18}$ s~\cite{Poulin2016}, lies one decade below
the latter. Right: energy-integrated incident angular distributions for
decay and annihilation in the same smooth NFW halo, divided by their
respective sky averages. Nuclear thresholds and detector acceptance have
not been applied to the right panel.}
\label{fig:source}
\end{figure*}

The xenon recoil lies near a nuclear diffraction minimum. Shifting
$c_A$ by $\pm0.2$ fm at fixed masses changes the required
$\Phi_G\sig$ to $(0.50$--$7.3)\times10^{-34}\,\mathrm{s^{-1}}$, while
$R_{\rm low}$ remains between $4.2\times10^{-4}$ and $1.1\times10^{-3}$
and the accepted mode moves in the range 245--251 keV. The thresholds and
endpoints are fixed by kinematics; the form factor and efficiency only
decide where inside the window the accepted maximum lies. These are
sensitivity tests, not a calibrated nuclear uncertainty. Where the
coherent response is this small, the $^{129}$Xe and $^{131}$Xe levels at
40 and 80 keV open within one line width above threshold and may compete
with it; more complete responses deserve
attention~\cite{Anand,Khan2026,LZth260905291}.
A finite scalar mediator also multiplies Eq.~\eqref{eq:dsigma} by
$[m_\phi^2/(m_\phi^2+2M_A\ER)]^2$; for $m_\phi=0.3$ GeV, above the
splitting $m_2-m_1$ so that the two-body decay $\chi_2\to\chi_1\phi$
stays closed (Section~\ref{sec:consistency}), the lifetime at fixed
$\sig$ is about 0.36 times the contact-limit value.
Figure~\ref{fig:source} shows the normalization relation and the angular
source dependence.

\section{Consistency conditions and target tests}
\label{sec:consistency}
For $f_X=1$ and $\sig=10^{-36}\,\mathrm{cm^2}$,
$t_U/\tau_X\simeq1.0\times10^{-3}$, taking
$t_U=13.8$ Gyr~\cite{Planck2018}, so the dominant parent density changes
little over this time. Equation~\eqref{eq:normalization} requires a
shorter lifetime if the allowed portal cross section is smaller, but not
indefinitely: while relativistic the daughters act as dark radiation, and
the CMB limit on decays into dark radiation,
$\tau\gtrsim5\times10^{18}$ s~\cite{Poulin2016}, translates into
$\sig\gtrsim10^{-38}\,\mathrm{cm^2}$ at $f_X=1$; the daughters become
nonrelativistic after a redshift factor of about 25, so the exact limit
is somewhat weaker. Constraints on other decay
topologies~\cite{Mau2022,Jho2025KM3,Rott2015} cannot be imported without
their energy-injection and propagation assumptions. The primordial
$\chi_1$ abundance and the parent abundance are independent inputs in
this effective description; $\chi_1$ need not be a thermal relic, and if
it were, its mass would be subject to the BBN and $N_{\rm eff}$ bounds on
MeV-scale thermal dark sectors~\cite{Sabti2020}. Their production must be
specified in a thermal or nonthermal completion.

The injection energy must also remain near the threshold: localization
requires $|E_0-\Eth|\lesssim\sigma_E$, i.e. $m_X$ aligned with $2\Eth$
to $\delta m_X/m_X\lesssim10^{-3}$. At fixed $(m_1,m_2)$, increasing
$E_0-\Eth$ to 0.2 MeV ($\delta m_X/m_X\simeq8\times10^{-4}$) gives a
recoil mode of 243.7 keV and $R_{\rm low}=0.0061$; at 0.5 MeV these
become 235.5 keV and 0.062. This alignment is shared by every
near-threshold interpretation of the event~\cite{LZth260906890}; we do
not propose a symmetry that enforces it. It is not the only viable
spectral arrangement; heavier daughters can also produce a peak near
248 keV after retuning the threshold and recomputing the velocity
distribution.

\textit{Elastic leakage.} An additional scalar elastic operator with the
same reference-cross-section convention gives
\begin{equation}
 \frac{N_{11}(14\text{--}225\,\keV)}{N_W}
 \simeq8.2\times10^3\frac{\sigma_{11}}{\sig}
 \label{eq:elastic}
\end{equation}
for this flux. Keeping the extra contribution below 0.1 event at $N_W=1$
requires $\sigma_{11}/\sig\lesssim1.2\times10^{-5}$, corresponding to an
amplitude ratio below $3.5\times10^{-3}$. This is a stated design target,
not a confidence limit. A common stabilizing $Z_2$ under which both
$\chi_1$ and $\chi_2$ are odd does not by itself forbid the diagonal
operator. A parity under which $\phi$ and one of the dark states are odd
does forbid it, but it also forbids the $\phi$--nucleon coupling and the
quartic in Eq.~\eqref{eq:invisible}, so the elastic suppression, the
portal coupling and the invisible width are set by the same symmetry
breaking; their tree-level coefficients and loop matching must be
controlled in a mediator completion.

\textit{The final dark state.} The recoil remains a single nuclear
interaction if $\chi_2$ escapes or decays invisibly. An explicit scalar
option is
\begin{equation}
 \mathcal L\supset-\frac{\lambda_{2111}}{3!}\chi_2\chi_1^3,
 \qquad \chi_2\to3\chi_1,
 \label{eq:invisible}
\end{equation}
which is open for Table~\ref{tab:bench}. Including the identical-particle
factor and exact three-body phase space, $\lambda_{2111}=10^{-6}$ gives
$\Gamma_{3\chi}=2.49\times10^{-18}$ GeV and $\tau_2=2.64\times10^{-7}$ s.
This supplies an invisible channel; its branching fraction must dominate
any visible channel in the chosen portal. For a mediator lighter than
$m_2-m_1\simeq243$ MeV, the two-body decay $\chi_2\to\chi_1\phi$ is
instead open through the same coupling that generates
Eq.~\eqref{eq:effective} and is prompt; such a light mediator requires
$\phi\to\chi_1\chi_1$ to dominate its Standard-Model modes, i.e. the
coupling hierarchy $g_{12}\gg g_{11}\gg g_N$ of
Ref.~\cite{LZth260906890}, with $g_{11}$ still subject to
Eq.~\eqref{eq:elastic}.
For a scalar mediator, a possible photon operator is
$-C_\gamma\phi F_{\mu\nu}F^{\mu\nu}/4$. Its coefficient is not fixed by
$C_N$, so the recoil normalization alone does not predict a photon signal.
Visible-decay constraints require the specified portal and spectrum
~\cite{Essig2013,Park2013,Huh2008,Jho2025halo}.

\textit{Other nuclei.} Define the fraction of incident particles above a
target's threshold,
\begin{equation}
 P_A=\int_{\Eth(M_A)}^\infty\dd E_1\,\mathcal F_G(E_1).
 \label{eq:gate}
\end{equation}
For the scalar benchmark flux, representative single-isotope values are
$P_{^{28}{\rm Si}}=0$ within the adopted velocity support,
$P_{^{40}{\rm Ar}}=1.9\times10^{-5}$,
$P_{^{72}{\rm Ge}}=0.10$, $P_{^{131}{\rm Xe}}=0.50$, and
$P_{^{184}{\rm W}}=0.67$. These fractions are not event-rate ratios;
the interaction, form factors, exposure and detector response still
enter. They illustrate how the physical line width turns an idealized
target threshold into a quantitative selection. Extended-window xenon
data and heavier targets provide direct tests, while the argon prediction
differs sharply from a cold exothermic line~\cite{LZth260915782}.

Cosmological daughters redshift according to $p(z)=p_0/(1+z)$ and reach
a target only from $1+z\leq p_0/p_{\rm th}(M_A)$, i.e. $z\lesssim10^{-3}$
near the Xe threshold; the dark-matter column within that volume is below
$10^{-2}\Dbar$ even with the Local Group included, so the Galactic
normalization suffices. For this benchmark the endothermic channel is
closed or strongly suppressed on the light nuclei of the overburden
(Eq.~\eqref{eq:gate}), and the elastic cross section allowed by
Eq.~\eqref{eq:elastic} gives an overburden optical depth of order
$10^{-10}$, so attenuation and terrestrial regeneration are negligible.

\section{Conclusions}
\label{sec:conclusions}
A long-lived dark parent provides both the energy and a calculable flux
for endothermic nuclear scattering. Its contribution scales with the
Galactic $D$ factor and $f_X/\tau_X$, separating the present source rate
from a reference annihilation rate. We have included the parent and
observer Doppler shifts and found a light-daughter benchmark whose
efficiency-weighted xenon spectrum remains concentrated near 248 keV.
The normalization to one event fixes a lifetime--cross-section relation,
Eq.~\eqref{eq:normalization}, whose coefficient the nuclear response near
the diffraction minimum can shift by a factor of 0.3--5, and the same
incident distribution predicts target-dependent threshold fractions.
These results give a quantitative decay source for a boosted endothermic
interpretation. The small elastic-channel requirement, nuclear-response
dependence and portal-specific constraints delimit the next steps toward
a complete test of the scenario.

\section*{Acknowledgements}
This work was supported by National Research Foundation of Korea (NRF)
grants funded by the Korean government (MSIT), Nos.~RS-2024-00340153 and
RS-2026-25607498.


\appendix
\section{Dirac realization and interaction dependence}
\label{app:dirac}
\setcounter{figure}{0}
\renewcommand{\thefigure}{A.\arabic{figure}}
\renewcommand{\theHfigure}{A.\arabic{figure}}

The recoil endpoints in Eqs.~\eqref{eq:endpoints}--\eqref{eq:threshold}
depend only on masses and incident energy. They are identical for scalar
and Dirac states at fixed $(m_1,m_2,E_1)$. Here we collect an alternative
Dirac realization, separating its interaction-dependent normalization
from this common kinematics.

Let $X\to\psi_1\bar\psi_1$ proceed through $-y_X X\bar\psi_1\psi_1$.
For this scalar Yukawa coupling,
$\Gamma_X=y_X^2m_X(1-4m_1^2/m_X^2)^{3/2}/(8\pi)$.
The flux in Eq.~\eqref{eq:flux} counts particles and antiparticles
together, assuming equal nuclear cross sections. A scalar-current portal is
\begin{equation}
 \mathcal L_D=\sum_q\frac{m_q}{\Lambda_S^3}
 (\bar\psi_2\psi_1)(\bar q q)+\mathrm{h.c.},\qquad
 G_n=\frac{f_n m_n}{\Lambda_S^3}.
 \label{eq:diracop}
\end{equation}
We take $f_p=f_n=0.30$ for universal quark coefficients with heavy-quark
matching~\cite{Hoferichter2017}; a portal restricted to $u,d,s$ requires
different matching. In the coherent one-body approximation $G_A=A G_n$,
\begin{align}
 \frac{\dd\sigma_A^D}{\dd\ER}
 =&\ \frac{G_A^2 M_A}{8\pi p_1^2}
 \big[(m_1+m_2)^2+2M_A\ER\big]\nonumber\\
 &\times\left(1+\frac{\ER}{2M_A}\right)F_A^2(\ER).
 \label{eq:diracdsigma}
\end{align}
Defining $\sigma_D=G_n^2\mu_n^2/\pi$, its relation to the scalar response is
\begin{equation}
 \frac{(\dd\sigma_A^D/\dd\ER)/\sigma_D}
 {(\dd\sigma_A/\dd\ER)/\sig}
 =\frac{(m_1+m_2)^2+2M_A\ER}{4m_1^2}.
 \label{eq:spinratio}
\end{equation}
This changes the weights within the allowed recoil interval, not its
endpoints. In particular, for nearly degenerate nonrelativistic states
the ratio approaches unity.

\begin{figure*}[t]
\centering\includegraphics[width=\textwidth]{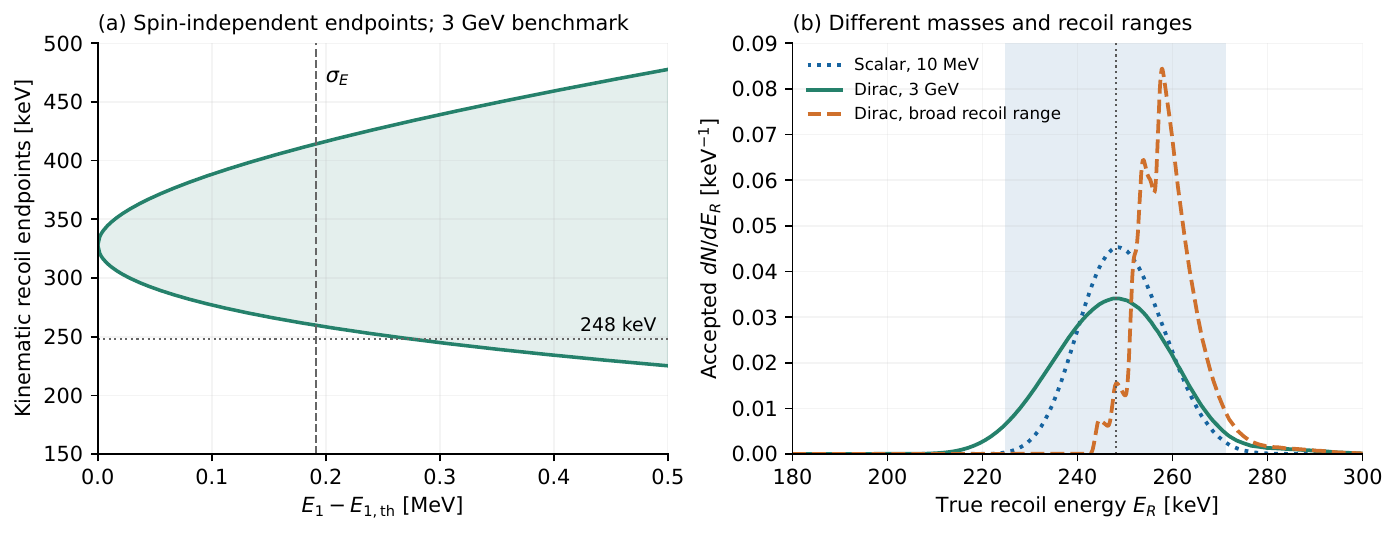}
\caption{Left: spin-independent recoil endpoints for the 3 GeV mass
choice in Eq.~\eqref{eq:diracbench} and the reference Xe nucleus. Scalar
and Dirac states with these masses give the same curves. The vertical
line marks one source energy standard deviation above threshold.
Right: the scalar 10 MeV and Dirac 3 GeV examples, normalized to one
event in $225$--$271$ keV using the same Galactic-source prescription,
natural Xe and LZ efficiency. Their differences do not isolate a spin
effect. Detector energy migration is not included. We also show the (Dirac) broad recoil range case deviated from the threshold with the parameters $m_1 = 6.0415$ GeV, $m_X = 18.0193$ GeV, $m_2 = 6.3052$ GeV for comparison.}
\label{fig:dirac}
\end{figure*}

For an illustrative GeV-scale example, choose
\begin{align}
 m_1&=3\,\GeV,&m_2&=3.0136698\,\GeV,\nonumber\\
 E_0&=3.0140059\,\GeV,&m_X&=6.0280118\,\GeV.
 \label{eq:diracbench}
\end{align}
The reference threshold recoil is 328 keV and $v_1=0.0963c$.
After Doppler broadening, isotope summation and efficiency weighting,
the mode is 248.3 keV, $\sigma_E=0.192$ MeV and $R_{\rm low}=0.033$.
The normalization is
\begin{equation}
 \Phi_G\sigma_D=3.62\times10^{-35}\,\mathrm{s^{-1}}
 \qquad (N_W=1).
 \label{eq:diracnorm}
\end{equation}
For $\tau_X=10^{19}$ s and $f_X=1$, this gives
$\Phi_G=534\,\mathrm{cm^{-2}s^{-1}}$ and $\Lambda_S=20.5$ GeV.
At these same masses, Eq.~\eqref{eq:spinratio} is 1.006 at 248 keV and
varies by only $0.031\%$ across 225--271 keV for the reference nucleus.
The normalized scalar and Dirac spectra are therefore almost identical.
For this Dirac flux, the elastic coefficient analogous to
Eq.~\eqref{eq:elastic} is $2.65\times10^4$, giving the illustrative
requirement $\sigma_{11}^D/\sigma_D\lesssim3.8\times10^{-6}$ for fewer
than 0.1 additional low-energy events.

Chemical equilibrium between the two states,
$n_2^{\rm eq}/n_1^{\rm eq}=(m_2/m_1)^{3/2}e^{-(m_2-m_1)/T}$, gives 0.90
for Eq.~\eqref{eq:diracbench} at $T=m_1/25$; a relic-density calculation
would in addition require the annihilation and conversion rates, QCD
inputs and the parent production~\cite{GriestSeckel,Garny2017}, and
collider viability requires a mediator
completion~\cite{ATLAS2015Monojet}. Finally, the invisible scalar quartic
in Eq.~\eqref{eq:invisible} does not apply to Dirac fields, and at the
masses in Eq.~\eqref{eq:diracbench} $m_2<3m_1$ closes a three-daughter
decay for either spin; the excited state must escape or have another
specified invisible channel.

\end{document}